\documentclass[sigconf]{acmart}

\microtypesetup{expansion=false,protrusion=false}

\newif\ifmarked
\markedfalse

\usepackage{booktabs}
\providecommand{\Description}[1]{}
\usepackage{listings}
\usepackage{multirow}
\usepackage{array}
\usepackage{xurl}
\usepackage[ruled,vlined,linesnumbered]{algorithm2e}
\SetAlCapSkip{2pt}      % space between caption and algorithm body
\SetAlgoSkip{smallskip}

\newcommand{\fullDist}{\texttt{FULL\_DISTANCE}\allowbreak}
\newcommand{\emptyDist}{\texttt{EMPTY\_DISTANCE}\allowbreak}

\acmConference[ICCA 2026]{4th International Conference on Computing
Advancements}{October 15--16, 2026}{Dhaka, Bangladesh}
\acmYear{2026}
\copyrightyear{2026}
\setcopyright{acmlicensed}
\acmISBN{979-8-XXXX-XXXX-X/26/10}
\acmDOI{10.1145/XXXXXXX.XXXXXXX}



\title[CleanCity-BinSense]{CleanCity-BinSense: An IoT-Enabled Smart
Waste Management System with Configurable Real-Time Fill Monitoring
and Nearest-Neighbor Route Optimization}

\author{Mohammad Adnan Kabir}
\authornote{Both authors contributed equally to this research.}
\email{adnankabir@iut-dhaka.edu}
\affiliation{%
  \institution{Islamic University of Technology}
  \city{Gazipur}
  \country{Bangladesh}
}

\author{Intifad Muhammad Sayeed}
\email{intifadsayeed@gmail.com}
\affiliation{%
  \institution{Jahangirnagar University}
  \city{Dhaka}
  \country{Bangladesh}
}

\usepackage{enumitem}
\setlist{itemsep=1pt, topsep=2pt, parsep=0pt, partopsep=0pt}
\begin{document}

% ================================================================
% ABSTRACT
% ================================================================
\begin{abstract}
Urban waste management in developing cities is often hindered by
inefficient fixed-schedule collection, overflowing waste bins, and
unnecessary fuel consumption caused by non-optimized collection routes.
This paper presents \textit{CleanCity-BinSense}, a low-cost,
end-to-end IoT-enabled smart waste management system designed to support
scalable real-time waste monitoring and demand-driven collection. The
proposed system integrates a solar-powered sensor node equipped with an
ultrasonic sensor for real-time bin fill-level monitoring. A key
contribution of the system is a configurable sensing model based on two
calibration parameters, \texttt{FULL\_DISTANCE} and
\texttt{EMPTY\_DISTANCE}, enabling deployment across bins of varying
sizes and geometries without requiring firmware modification. Fill
percentage is computed using a geometry-configurable linear
normalization algorithm validated through hardware experiments with a
mean absolute error (MAE) of 0.38~cm, remaining within the
manufacturer-specified tolerance of the sensor hardware. Sensor readings
are transmitted via Wi-Fi to a centralized web platform that provides
role-based dashboards for administrators, operators, and drivers, along
with a public real-time bin-status map. The system further incorporates
a lightweight nearest-neighbor route planning algorithm using SQL
Server's spatial function to generate proximity-based collection routes
with low computational overhead. Experimental evaluation demonstrates an
average end-to-end system latency of 5.3~seconds, dominated primarily by
the configurable sensing interval rather than network overhead, while
route generation for typical urban collection zones completes in under
100~ms. These results demonstrate the feasibility of deploying a
low-cost, configurable, and infrastructure-light smart waste management
system capable of supporting heterogeneous urban waste networks in
resource-constrained environments such as Dhaka, Bangladesh.
\end{abstract}

\maketitle

% ================================================================
\section{Introduction}
% ================================================================

Rapid urbanization in developing countries has intensified the challenges
of municipal solid waste (MSW) management. In cities such as Dhaka,
Bangladesh, one of the most densely populated cities in the world,
waste collection vehicles follow fixed schedules largely independent of
actual bin occupancy~\cite{gutierrez2015smart}. This creates two
major inefficiencies: bins overflow before collection arrives,
creating public health hazards, while vehicles are dispatched to largely
empty bins, wasting fuel and driver time.

The Internet of Things (IoT) paradigm offers a compelling solution. By
embedding sensors in bins and connecting them to a centralized platform,
city authorities can shift from \textit{time-based} to
\textit{demand-based} collection. Previous studies have demonstrated that such
systems can reduce collection costs by 20--40\% in pilot
deployments~\cite{longhi2012waste, anagnostopoulos2017challenges}.
However, most existing systems share a critical practical limitation:
sensing ranges are \textit{hardcoded} for a specific bin geometry,
limiting deployment across heterogeneous bin infrastructures without
modifying and re-flashing firmware. This severely limits real-world
scalability where bins of varying heights and capacities coexist on the
same city network.

This paper presents \textbf{CleanCity-BinSense}, an integrated
IoT-enabled waste management system that addresses these limitations
while improving operational efficiency and scalability. The key
contributions are:

\begin{itemize}
  \item A solar-powered IoT sensor node in a custom 3D-printed
  enclosure with a \textbf{configurable sensing range}, operators
  set two firmware parameters, \texttt{FULL\_DISTANCE} (the sensor
  reading when the bin is at full capacity) and \texttt{EMPTY\_DISTANCE}
  (the reading when the bin is empty) per bin to match any bin
  geometry, enabling uniform deployment across heterogeneous bin
  infrastructure

  \item A calibrated, geometry-independent linear fill-percentage
  algorithm validated with real hardware (MAE 0.38~cm).

  \item A full-stack web platform with role-based dashboards (Admin,
  Operator, Driver) and a public bin map, operating without mandatory
  cloud dependency.

  \item An automated nearest-neighbor route optimization algorithm
  using SQL Server's \texttt{STDistance()} geodesic distance function
  for accurate proximity-based stop ordering, with sub-100~ms
  computation time enabling real-time responsive planning.

  \item A complete, physically built and tested prototype demonstrating
  end-to-end integration from sensor to optimized driver route.
\end{itemize}

The remainder of this paper is organized as follows. Sections 2--6
present the related work, system design, implementation, and
experimental evaluation, while Sections 7 and 8 discuss limitations and
conclude the paper.

% ================================================================
\section{Background \& Prior Research}
\label{sec:related}
% ================================================================

\textbf{Sensor-Based Fill Detection:}
Gutierrez et al.~\cite{gutierrez2015smart} proposed one of the earliest
IoT waste bin systems using infrared sensors and GSM data transmission.
While effective for proof-of-concept, GSM introduces recurring
operational costs that are unsuitable for large-scale city deployments.
Khan et al.~\cite{khan2024efficient} proposed a knapsack-based
IoT-assisted waste collection system that prioritizes bins by toxicity
and fill level, demonstrating reduced collection visits in simulation,
but without per-bin sensing configuration. None of these systems expose a
configurable sensing range, a critical limitation where bins of
varying heights coexist on the same city network.
CleanCity-BinSense addresses this gap by making the sensing range fully
configurable per bin and integrating it with a complete role-based
management web platform.

\textbf{Advanced Optimization Techniques:}
Jerbi et al.~\cite{jerbi2025dbeso} proposed a sophisticated
optimization framework that combines IoT sensors with a Dynamic Bald Eagle
Search Optimization algorithm (DBESO) and a Kernel-based Extreme
Learning Machine (KELM) for accurate waste status prediction. Although
their approach delivers strong predictive performance in simulation, it
is evaluated without physical hardware deployment and requires
specialized machine learning expertise and computational infrastructure
that may be impractical for resource-constrained city
environments~\cite{ahmed2024ai}. Ahmed et al.~\cite{ahmed2024ai}
conducted a comprehensive review of AI and IoT architectures for
municipality waste management, identifying real-time sensor integration,
role-based access, and demand-based scheduling as the most impactful
practical requirements for developing city deployments.
CleanCity-BinSense directly targets these requirements with a
lightweight, physically deployed architecture.

\textbf{Route Optimization:}
Anagnostopoulos et al.~\cite{anagnostopoulos2017challenges}
surveyed
a dynamic routing algorithm based on real-time fill data, demonstrating
significant travel distance reductions. Longhi et
al.~\cite{longhi2012waste} integrated fill data with external GIS
software for route planning. Addas et al.~\cite{addas2024waste} deployed
an IoT and cloud analytics system across 10 locations in Lahore with
dynamic route optimization, achieving measurable improvements in route
efficiency and fuel consumption. However, these approaches rely on
external GIS infrastructure~\cite{longhi2012waste,
gutierrez2015smart} or third-party cloud analytics
platforms~\cite{addas2024waste}, introducing additional infrastructure
dependencies that increase deployment complexity and cost.
CleanCity-BinSense embeds proximity-based route optimization directly
into the operator web platform using SQL Server spatial functions,
requiring no external GIS infrastructure or cloud dependency.

\textbf{Platform and Dashboard Design:}
Folianto et al.~\cite{folianto2015smartbin} presented SmartBin for
Singapore, focusing on sensor hardware and a wireless mesh network;
their system does not provide role-differentiated access for operators
and drivers, nor route optimization capability. Chowdhury
et al.~\cite{chowdhury2020iot} specifically targeted Dhaka using
NodeMCU and the ThingSpeak cloud platform for data storage and
visualization. However, ThingSpeak dependency means that the system requires
continuous internet connectivity and cannot operate if the cloud
service is unavailable, a significant risk in environments with
unreliable internet infrastructure. CleanCity-BinSense operates on a
locally-hosted server, requiring only a site-local network between bins
and the server, and provides role-based dashboards absent from both
prior systems.

\textbf{Solar-Powered IoT:}
Solar-powered smart waste monitoring has been explored
as a practical direction for smart city planning~\cite{kabir2020solar}.
Our system integrates solar power into the bin sensor unit for
sustainable autonomous operation across diverse deployment sites.

Table~\ref{tab:comparison} compares CleanCity-BinSense with key prior
works across six practical deployment dimensions.

\begin{table}[h]
\caption{Comparison with related smart waste management systems.
HW~=~hardware prototype, Dash~=~web dashboard, Route~=~route
optimization, Cfg.~Range~=~configurable sensing range per bin.}
\label{tab:comparison}
\footnotesize
\begin{tabular}{lcccccc}
\toprule
\textbf{System} & \textbf{HW} & \textbf{Solar} & \textbf{Dash}
  & \textbf{Roles} & \textbf{Route} & \textbf{Cfg.} \\
\midrule
Gutierrez et al.~\cite{gutierrez2015smart}  & \checkmark & $\times$ & $\times$ & $\times$ & $\times$ & $\times$ \\
Folianto et al.~\cite{folianto2015smartbin} & \checkmark & $\times$ & \checkmark & $\times$ & $\times$ & $\times$ \\
Chowdhury et al.~\cite{chowdhury2020iot}    & \checkmark & $\times$ & \checkmark & $\times$ & $\times$ & $\times$ \\
Khan et al.~\cite{khan2024efficient}        & $\times$   & $\times$ & $\times$ & $\times$ & \checkmark & $\times$ \\
Jerbi et al.~\cite{jerbi2025dbeso}          & $\times$   & $\times$ & $\times$ & $\times$ & \checkmark & $\times$ \\
Addas et al.~\cite{addas2024waste}          & \checkmark & $\times$ & $\times$ & $\times$ & \checkmark & $\times$ \\
\textbf{CleanCity-BinSense}                 & \checkmark & \checkmark & \checkmark & \checkmark & \checkmark & \checkmark \\
\bottomrule
\end{tabular}
\end{table}

Unlike previous studies that focus on individual aspects such as
fill-level detection, route optimization, or cloud-based monitoring,
\textbf{CleanCity-BinSense} integrates these capabilities into a single
deployable smart waste management platform. The system introduces
several practical innovations:

\begin{itemize}
    \item \textbf{Configurable per-bin sensing range} for bins of varying dimensions.
    \item \textbf{Role-based web dashboards} for administrators, operators, and drivers.
    \item \textbf{Built-in SQL Server spatial route optimization} without external GIS or cloud services.
    \item \textbf{Self-hosted, solar-powered architecture} for reliable and sustainable municipal deployment.
\end{itemize}

Together, these features provide a practical, scalable solution that
addresses several deployment limitations identified in existing smart
waste management systems.

% ================================================================
\section{System Architecture}
\label{sec:architecture}
% ================================================================

\begin{figure}[htbp]
\centering
\includegraphics[width=0.85\linewidth]{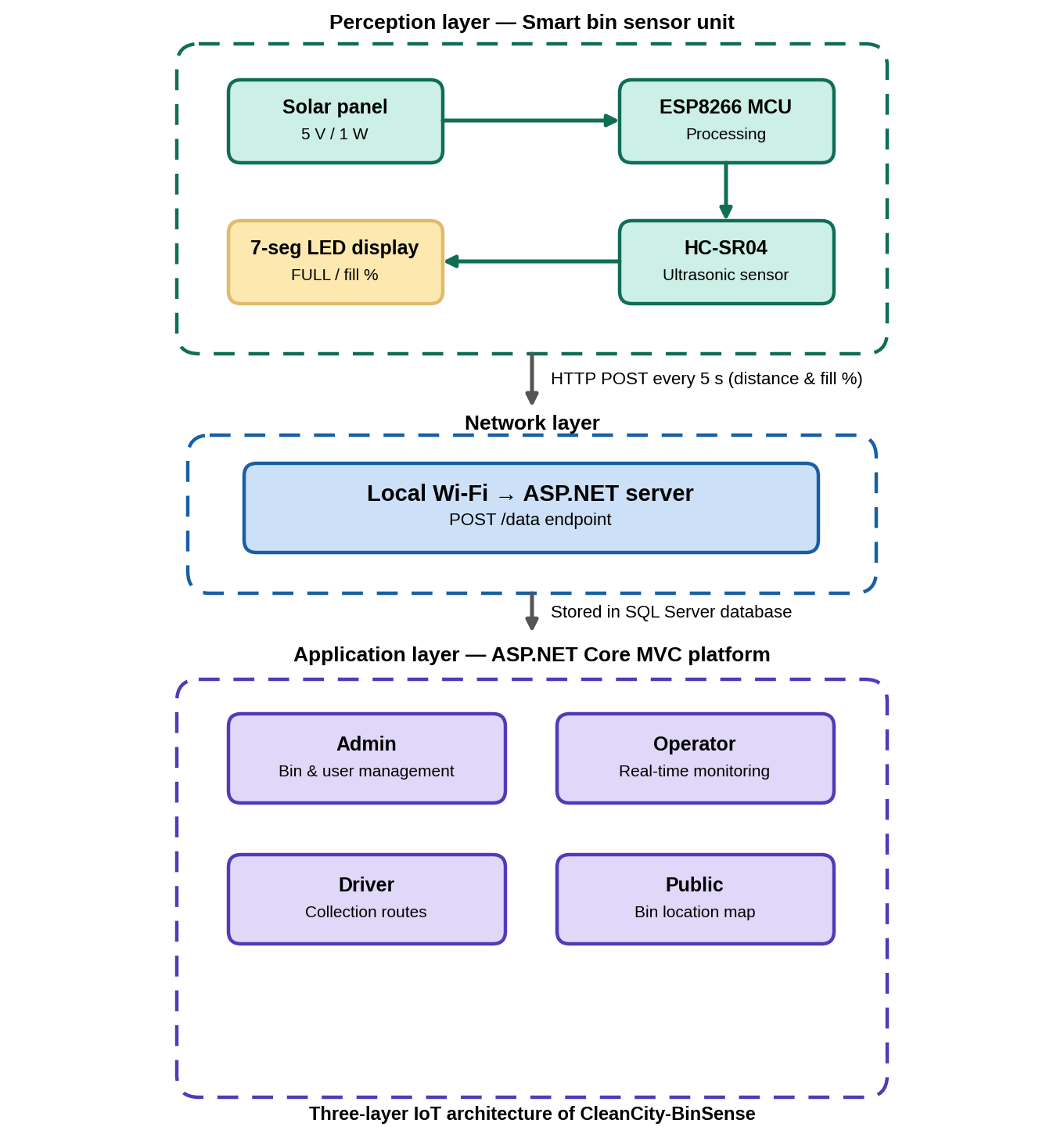}
\Description{Block diagram showing the three-layer IoT architecture of
CleanCity-BinSense. The perception layer contains a solar panel, an
ESP8266 microcontroller, an HC-SR04 ultrasonic sensor, and a 7-segment
LED display. The network layer shows a local Wi-Fi connection to an
ASP.NET server via HTTP POST. The application layer shows four
role-based modules: Admin, Operator, Driver, and Public.}
\caption{Three-layer IoT architecture of CleanCity-BinSense showing
perception, network, and application layers with the route optimization
module.}
\label{fig:architecture}
\end{figure}

CleanCity-BinSense follows the canonical three-layer IoT
architecture~\cite{sethi2017iot}, the most widely adopted
foundational model in IoT system design, as shown in
Figure~\ref{fig:architecture}: the \textit{Perception Layer} (physical
sensing), the \textit{Network Layer} (Wi-Fi communication), and the
\textit{Application Layer} (web platform and route optimization).

Each bin is registered in the system with its GPS coordinates, physical
capacity specifications, and its own calibrated \fullDist{} and
\emptyDist{} values. This per-bin configuration enables the web platform
to correctly interpret fill percentages from bins of different heights,
making the system deployable across heterogeneous urban bin
infrastructure without any hardware changes.

\textbf{Perception Layer:} Each smart bin is fitted with a sensor unit
at the lid. The HC-SR04 ultrasonic distance
sensor~\cite{hcsr04datasheet} emits a pulse and measures the echo
return time from the waste surface. The microcontroller processes this
signal, computes the fill percentage using its calibrated range
parameters, and displays the status on a 7-segment LED display.

\textbf{Network Layer:} The microcontroller connects to the site-local
Wi-Fi infrastructure (e.g., a municipal access point or premises
network) and transmits sensor readings to the web platform server via
HTTP POST every 5 seconds. The payload includes distance, fill
percentage, and bin identifier.

\textbf{Application Layer:} The web platform server stores readings in
SQL Server, serves role-specific dashboards (Admin, Operator, Driver),
runs route optimization on demand, and serves a public bin map. A
dedicated \texttt{/data} endpoint receives all hardware readings.

% ================================================================
\section{Hardware Design}
\label{sec:hardware}
% ================================================================

\subsection{Components}

The sensor unit comprises:
\begin{itemize}
 \item \textbf{ESP8266 Microcontroller:} Serves as the primary processing unit,
providing Wi-Fi connectivity, GPIO control, and embedded networking
capability for transmitting sensor readings to the web platform
\cite{esp8266datasheet,esp8266techref}.

  \item \textbf{HC-SR04 Ultrasonic Distance Sensor:}
  Measures the distance between the bin lid and the waste surface. The
  sensor operates within a range of 2--400~cm, with an approximate
  resolution of 0.3~cm and an operating frequency of 40~kHz. The
  effective sensing range is calibrated per bin according to its
  physical dimensions.

  \item \textbf{4-Digit 7-Segment LED Display:} Displays the current
  fill percentage locally, or ``FULL'' when the bin reaches capacity,
  without requiring network connectivity.

  \item \textbf{Solar Panel with Charging Circuit:} Provides autonomous
  power supply, eliminating the need for wired electrical
  infrastructure and enabling flexible deployment.

  \item \textbf{Custom 3D-Printed Enclosure:} A PLA-printed two-part
  enclosure designed with dedicated cutouts for the ultrasonic sensor,
  LED display, and cable routing.
\end{itemize}

\subsection{3D-Printed Enclosure Design}

\begin{figure}[h]
  \centering
  \includegraphics[width=0.32\columnwidth]{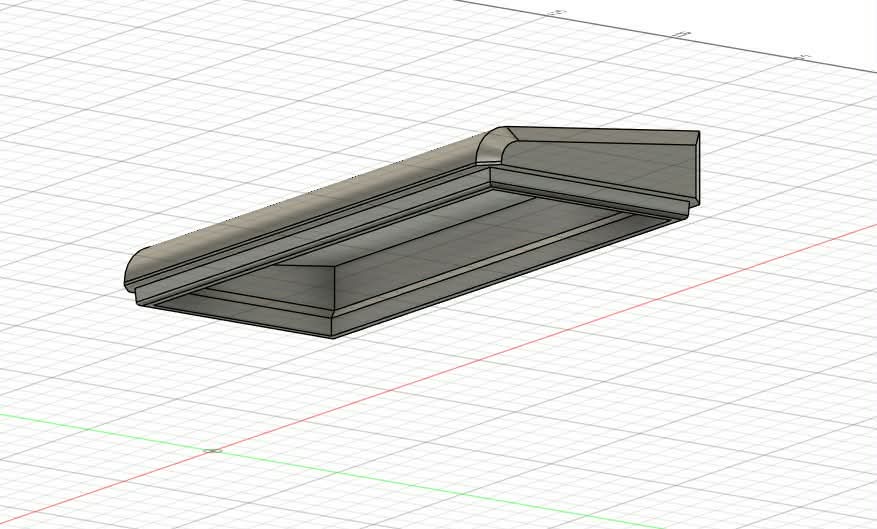}
  \hfill
  \includegraphics[width=0.32\columnwidth]{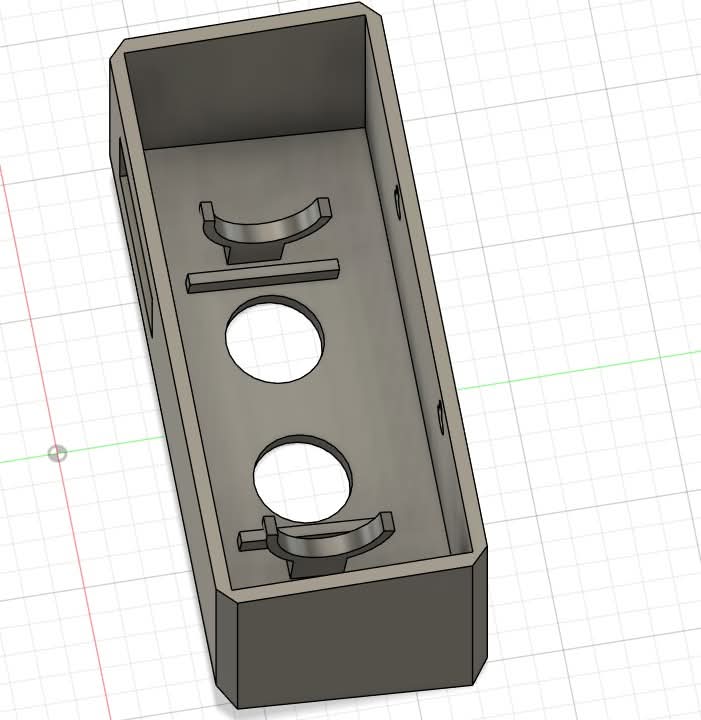}
  \hfill
  \includegraphics[width=0.32\columnwidth]{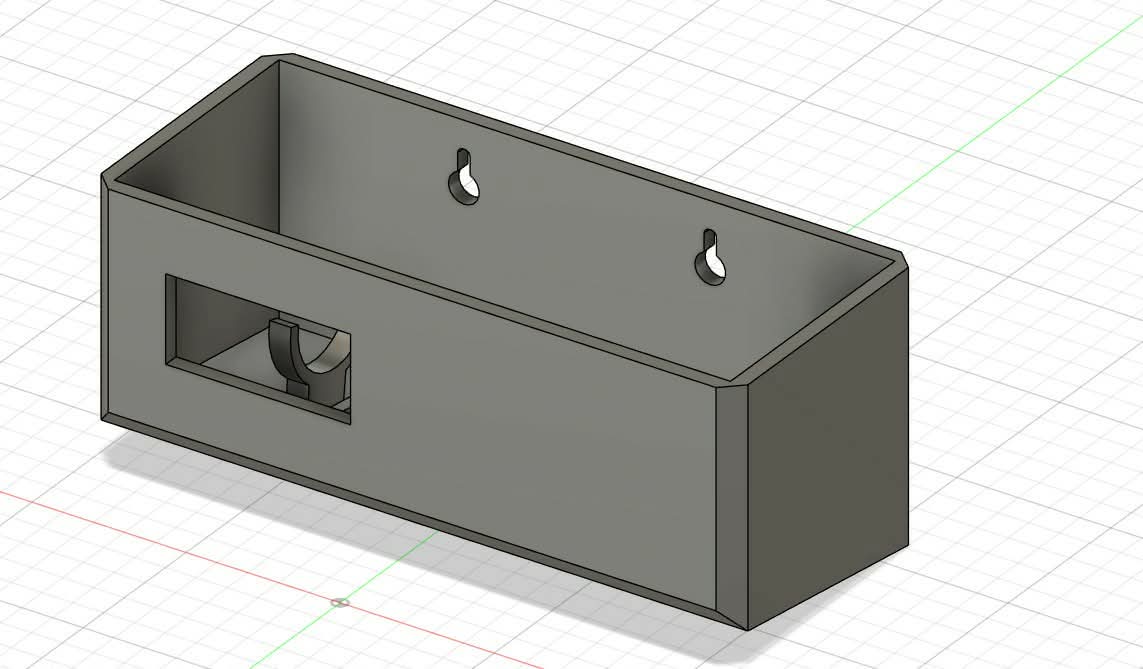}
  \Description{Three CAD renders of the 3D-printed sensor enclosure.
  Left: the lid showing a recessed sliding channel for tool-free
  assembly. Centre: top view of the main body showing two circular
  ports for ultrasonic transducers and internal PCB mounting clips.
  Right: rear view showing keyhole mounting slots and an internal
  cable management channel.}
  \caption{3D CAD design of the sensor enclosure: (a) lid with sliding
  channel, (b) main body with ultrasonic sensor ports and internal PCB
  clips, (c) rear view showing keyhole mounting slots and internal cable
  management channel.}
  \label{fig:cad}
\end{figure}

The enclosure was designed in CAD software and 3D printed in PLA.
Figure~\ref{fig:cad} shows the three-part CAD design: (a) the lid
featuring a recessed sliding channel for tool-free assembly, (b) the
main body with two circular ports for the ultrasonic transducers and
internal PCB mounting clips, and (c) the rear view with keyhole
mounting slots for wall or bin-lid installation.

\subsection{Assembled Prototype}

\begin{figure}[h]
  \centering
  \includegraphics[width=0.49\columnwidth]{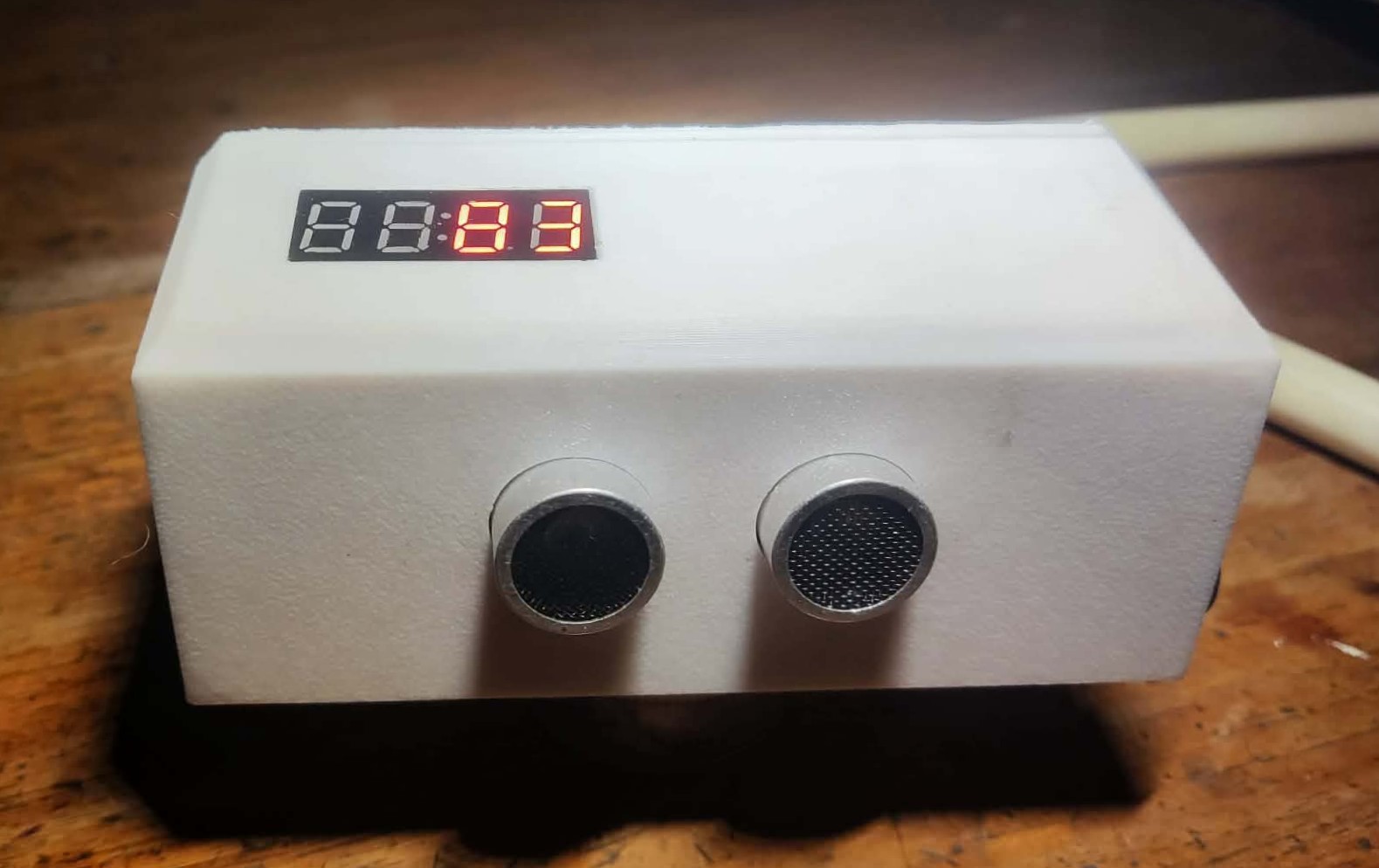}
  \hfill
  \includegraphics[width=0.49\columnwidth]{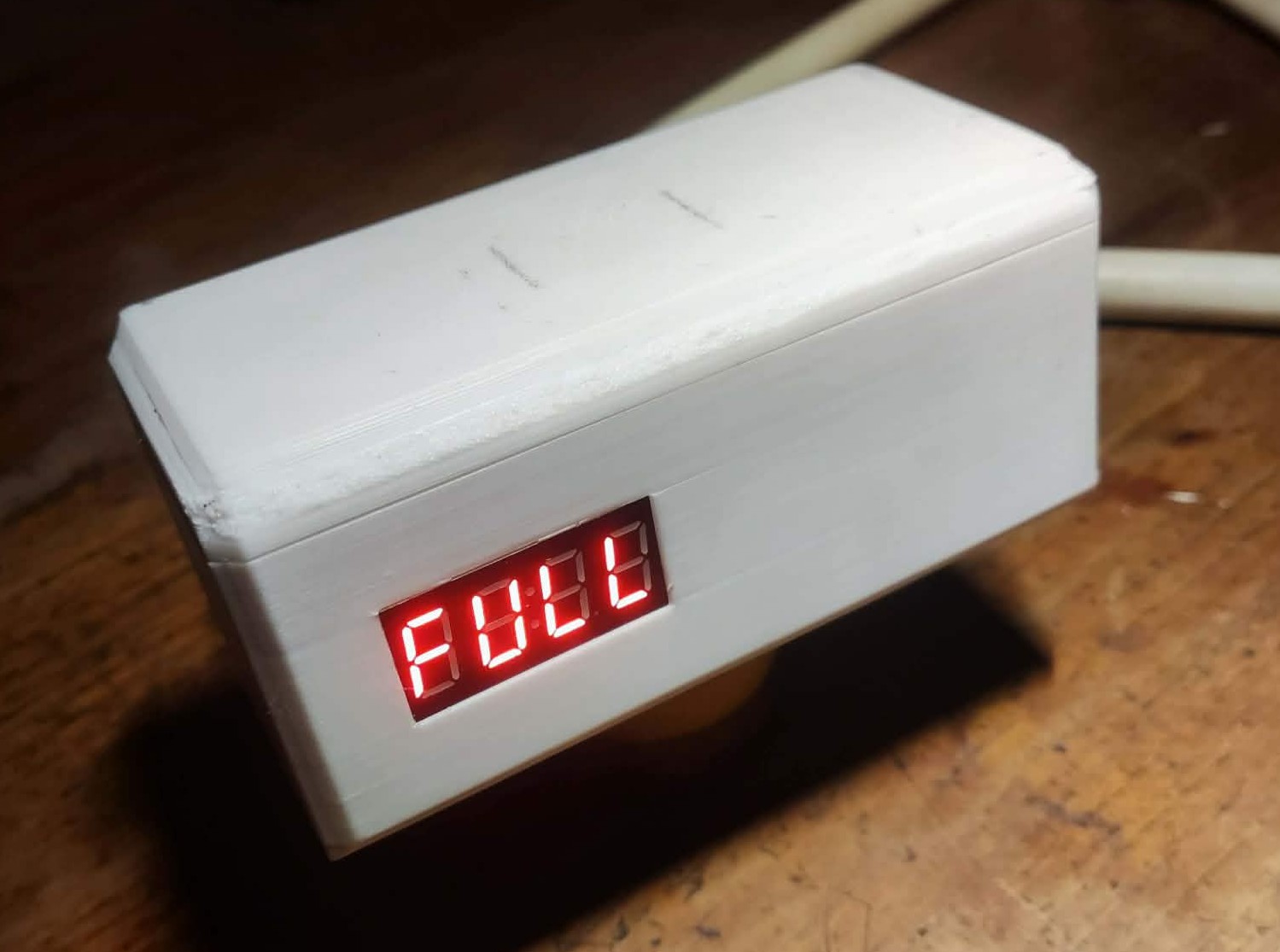}
  \Description{Two photographs of the assembled white 3D-printed
  prototype. Left: front view showing two circular ultrasonic sensor
  transducers and a 4-digit 7-segment LED display showing the current
  fill percentage. Right: the same unit displaying the word FULL in
  red LED segments when the bin reaches maximum capacity.}
  \caption{Assembled prototype: normal operation showing fill percentage
  (left); ``FULL'' alert when bin reaches capacity (right).}
  \label{fig:hardware}
\end{figure}

Figure~\ref{fig:hardware} shows the fully assembled prototype. The
front face exposes both ultrasonic transducers and the LED display
showing live fill percentage or ``FULL'' at maximum capacity.

\subsection{Configurable Fill Level Calculation}

A key design principle of CleanCity-BinSense is that the sensing range
is \textit{not hardcoded}. The firmware exposes two operator-configurable
parameters:

\begin{itemize}
  \item \texttt{FULL\_DISTANCE} ($D_{full}$): the distance reading (cm)
  when the bin is completely full. Set to match the bin's full-state
  geometry (e.g., waste level just below the sensor).
  \item \texttt{EMPTY\_DISTANCE} ($D_{empty}$): the distance reading
  (cm) when the bin is completely empty. Set to match the bin's
  internal depth.
\end{itemize}

These two values can be changed in the firmware configuration and
uploaded once per bin during installation, requiring no hardware
modifications. The same sensor unit can therefore serve a shallow
30~cm public litter bin or a deep 120~cm industrial waste container by
simply updating these two values.

The fill percentage $F$ is computed by the following
geometry-independent linear normalization formula:

\begin{equation}
F = \operatorname{constrain}\!\left(
  \frac{D_{empty} - d}{D_{empty} - D_{full}} \times 100,\ 0,\ 100
\right)
\label{eq:fill}
\end{equation}

\begin{sloppypar}
where $\operatorname{constrain}(x, 0, 100)$ clamps the result to
$[0, 100]$ to handle edge cases (e.g., sensor noise beyond calibrated
bounds).
\end{sloppypar}

\textbf{Worked example:} For a bin calibrated with $D_{full} = 4.0$~cm
and $D_{empty} = 73.0$~cm, a reading of $d = 16.4$~cm gives:
\begin{equation}
F = \frac{73.0 - 16.4}{73.0 - 4.0} \times 100
  = \frac{56.6}{69.0} \times 100 \approx 82\%
\end{equation}

This matches the live reading of 81\% observed during hardware testing
(within rounding tolerance), confirming the formula's validity.

\textbf{Boundary behavior:} When $d \leq D_{full}$, the formula returns
100\% (bin full). When $d \geq D_{empty}$, it returns 0\% (bin empty).
The \texttt{constrain} call ensures stable output in both cases
regardless of sensor noise.

\textbf{Firmware implementation.}
Algorithm~\ref{alg:sensing} shows the two firmware functions that
implement the sensing pipeline on the ESP8266. Only the two calibration
constants, \texttt{FULL\_DISTANCE}
($D_{full}$) and \texttt{EMPTY\_DISTANCE} ($D_{empty}$), differ
between bin deployments; the \texttt{getDistanceCM()} and
\texttt{getFillPercent()} functions are identical for every bin
regardless of its physical geometry. \texttt{getDistanceCM()} triggers
the HC-SR04 pulse sequence and converts the echo duration to
centimeters using the speed of sound ($0.0343$~cm/$\mu$s).
\texttt{getFillPercent()} applies the linear normalization of
Eq.~\ref{eq:fill} with \texttt{constrain} clamping, producing a
stable 0--100\% output for any calibrated range.

\begin{algorithm}[h]
\DontPrintSemicolon
\SetAlgoLined
\KwIn{Calibration constants $D_{full}$, $D_{empty}$ (set once per bin
      at installation)}
\KwOut{Distance reading $d$ (cm); Fill percentage $F$ (\%)}
\SetKwFunction{FGetDist}{getDistanceCM}
\SetKwFunction{FGetFill}{getFillPercent}
\SetKwProg{Fn}{Function}{:}{}

\Fn{\FGetDist{}}{
  Trigger ultrasonic pulse on \textit{TRIG\_PIN}\;
  \textit{duration} $\leftarrow$ echo pulse width on \textit{ECHO\_PIN}
    (timeout 30{,}000~$\mu$s)\;
  \If{\textup{duration} $= 0$}{
    \Return{$-1$} \tcp*{invalid reading}
  }
  \Return{$(\textit{duration} \times 0.0343) / 2$} \tcp*{cm, using speed of sound}
}

\Fn{\FGetFill{$d$}}{
  \If{$d \leq D_{full}$}{\Return{$100$}}
  \If{$d \geq D_{empty}$}{\Return{$0$}}
  $F \leftarrow \dfrac{D_{empty} - d}{D_{empty} - D_{full}} \times 100$\;
  \Return{$\operatorname{constrain}(F,\ 0,\ 100)$}
}
\caption{Distance Measurement and Fill-Level Computation (ESP8266
Firmware). Only $D_{full}$ and $D_{empty}$ differ between bin
deployments; both functions are identical for every bin geometry.}
\label{alg:sensing}
\end{algorithm}

The \textit{echo timeout} of 30{,}000~$\mu$s bounds the maximum
measurable distance to approximately 515~cm, safely exceeding the
calibrated $D_{empty}$ range used in this deployment. Changing only
$D_{full}$ and $D_{empty}$ adapts the same firmware to any bin
geometry, requiring no further code modifications.

\subsection{Data Transmission}

The ESP8266 posts sensor data every 5 seconds. The transmitted payload
consists of four fields: measured distance, computed fill percentage,
bin identifier (\textit{bin\_id}), and device MAC address
(\textit{device\_mac}), encoded as a URL-form HTTP POST request to the
server's \texttt{/data} endpoint.

Algorithm~\ref{alg:loop} shows the corresponding firmware transmission
loop. The sensor reading and fill computation are invoked every
5~seconds; the result is encoded as a URL-form payload and dispatched
via HTTP POST. Invalid readings ($d = -1$) are discarded before
transmission to prevent corrupt data reaching the server.

\begin{algorithm}[h]
\DontPrintSemicolon
\SetAlgoLined
\KwIn{Wi-Fi connection status; calibration constants $D_{full}$,
      $D_{empty}$; server URL}
\KwOut{Periodic HTTP POST transmission of sensor readings}
\SetKwFunction{FGetDist}{getDistanceCM}
\SetKwFunction{FGetFill}{getFillPercent}
\While{\textup{true}}{
  \If{\textup{Wi-Fi status} $=$ CONNECTED}{
    $d \leftarrow$ \FGetDist{}\;
    $F \leftarrow$ \FGetFill{$d$}\;
    \If{$d > 0$}{
      \textit{payload} $\leftarrow$ \{distance: $d$, fill: $F$\}\;
      Send HTTP POST \textit{payload} to server \texttt{/data}
        endpoint\;
    }
  }
  Wait 5 seconds\;
}
\caption{Main Firmware Loop: Sensor Polling and HTTP POST
Transmission}
\label{alg:loop}
\end{algorithm}

The \textit{bin\_id} uniquely identifies the bin; \textit{device\_mac}
allows the server to match readings to registered hardware. If a
transmission fails, the firmware retries every 5 seconds for up to 5
minutes before discarding the reading.

% ================================================================
\section{Software Platform}
\label{sec:software}
% ================================================================

\subsection{Technology Stack}

The backend is built on \textbf{ASP.NET Core MVC}
(.NET~9.0)~\cite{microsoft2024aspnetcore} with
\textbf{Entity Framework Core} for data
access~\cite{microsoft2024efcore} and \textbf{SQL Server}
for storage. The frontend uses \textbf{Razor Views} with Bootstrap for
responsive UI and \textbf{Leaflet.js}~\cite{agafonkin2010leaflet} for
interactive map visualization. REST API endpoints enable hardware
communication and future mobile app integration.

\subsection{Per-Bin Configuration Management}

When registering a new bin in the system, the Admin records the bin's
\texttt{FULL\_DISTANCE} and \texttt{EMPTY\_DISTANCE} values alongside
its GPS coordinates and capacity. The server stores these per-bin
parameters and uses them to validate incoming sensor readings and
display fill-level history correctly for bins of different geometries.
This design means that a city deploying CleanCity-BinSense across
hundreds of bins of varying sizes does not require any custom firmware
builds, only a one-time calibration entry at installation.

\subsection{Role-Based Access Control}

The web platform provides three differentiated user roles:

\begin{itemize}
  \item \textbf{Admin:} Responsible for route planning and system
  oversight. The Admin dashboard presents summary metrics, recent
  pickups, and route status information. Admin users can generate new
  collection routes by selecting a zone, date, truck, driver, and
  minimum fill threshold. The system then creates a route with ordered
  stops using the configured planning procedure.

  \item \textbf{Operator:} Responsible for bin and reading management.
  Operator users can register new bins, view all bins with their latest
  fill levels, and submit new sensor readings. The operator dashboard
  highlights bins that need attention, including bins marked as needing
  pickup or overflowing.

  \item \textbf{Driver:} Responsible for route execution in the field.
  Driver users view only the routes assigned to them, inspect route
  details on an interactive map, and record pickup completion for each
  stop. The interface is optimized for quick use during field
  operations and shows route progress clearly.
\end{itemize}

\subsection{Public Bin Map}

The platform includes a publicly accessible map that displays nearby
bins with live status information. The map is centered around the
user's location when available, or around a default city location
otherwise. Bins are retrieved through a dedicated API endpoint that
returns bins within a selected radius, along with their current fill
level, status, last reading time, and distance from the selected point.

Bins are color-coded by status to improve readability: green for normal
bins, orange for bins that need pickup, and red for overflowing bins.
Citizens can use this map to understand the current condition of bins
in their area without accessing the authenticated dashboards.

\begin{figure}[h]
  \centering
\includegraphics[width=0.88\columnwidth]{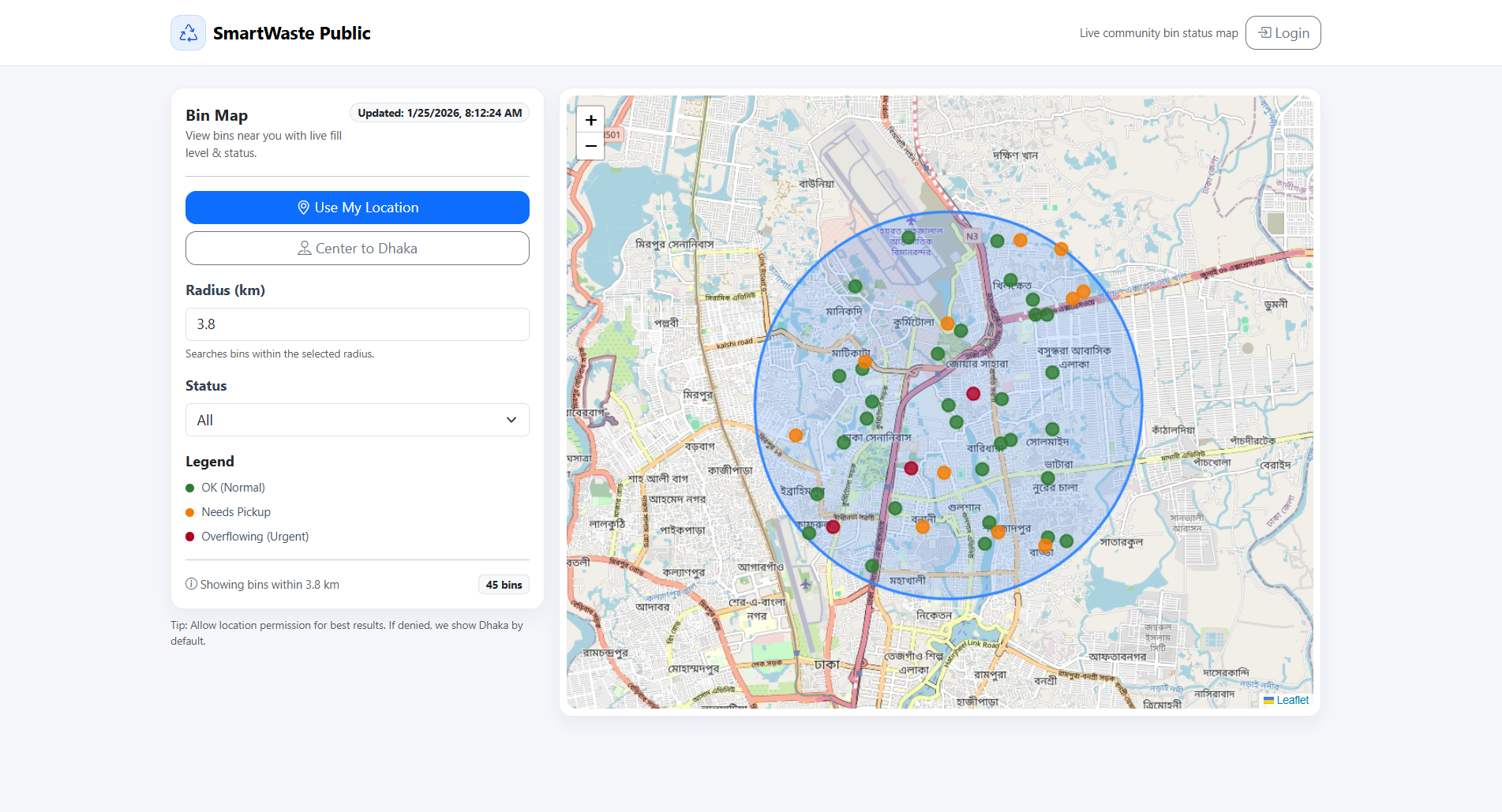}  \Description{Screenshot of the public SmartWaste bin map web
  interface showing a Leaflet map centered on Dhaka, Bangladesh.
  A circular radius overlay highlights the search area of 3.8 km.
  Bins are shown as color-coded circle markers: green for normal
  status, orange for needs pickup, and red for overflowing. A side
  panel shows filter controls for radius and status, and a legend
  explaining the color codes. The map shows 45 bins in the area.}
  \caption{Public bin map with color-coded fill status: green
  (Normal), orange (Needs Pickup), red (Overflowing/Urgent).}
  \label{fig:public_map}
\end{figure}

\subsection{Route Optimization Algorithm}

When an Admin creates a collection route, the system applies a
nearest-neighbor heuristic implemented in C\# within the ASP.NET Core
backend, supported by SQL Server's spatial
\texttt{geography\allowbreak::\allowbreak STDistance()} function for
accurate geodesic distance measurement~\cite{mssql2024stdistance}. The
algorithm is summarized in
Algorithm~\ref{alg:nn}. The nearest-neighbor heuristic is a classical
approximation approach for route ordering problems related to the
traveling salesman problem~\cite{rosenkrantz1977tsp}.

\begin{algorithm}[h]
\DontPrintSemicolon
\SetAlgoLined
\KwIn{Set $B$ of bins requiring collection (each with GPS coordinates);
      zone reference point $z$; minimum fill threshold $\tau$}
\KwOut{Ordered stop list $R$; estimated total route distance $D$}
Filter $B$: retain only bins where $\mathrm{fill} \geq \tau$\;
$\mathrm{current} \leftarrow z$\;
$R \leftarrow [\,]$\;
$D \leftarrow 0$\;
\While{$B \neq \emptyset$}{
  $b^{*} \leftarrow \displaystyle\arg\min_{b \in B}\;
    \mathrm{STDistance}(\mathrm{current},\, b)$\;
  Append $b^{*}$ to $R$\;
  $D \leftarrow D + \mathrm{STDistance}(\mathrm{current},\, b^{*})$\;
  $\mathrm{current} \leftarrow b^{*}$\;
  $B \leftarrow B \setminus \{b^{*}\}$\;
}
\Return{$R,\; D$}\;
\caption{Nearest-Neighbor Route Optimization with Spatial Distance}
\label{alg:nn}
\end{algorithm}

\textbf{Spatial distance measurement.} Inter-bin distances are computed
using SQL Server's \texttt{geography\allowbreak::\allowbreak
STDistance()} function, which calculates the geodesic (great-circle)
distance in meters between two WGS-84 latitude/longitude points on an
ellipsoidal Earth model. This is more accurate than planar Euclidean
distance for urban-scale routes and requires no external mapping API.
Each bin's coordinates are stored as \texttt{DECIMAL(9,6)}
latitude/longitude fields in the \texttt{Bins} table and queried as SQL
Server \texttt{geography} objects at route-planning time. The distance
from a reference point $p$ to a bin at coordinates $(\phi, \lambda)$ is
computed as:

\begin{lstlisting}[label={lst:stdist},
  caption={Geodesic distance computation using STDistance()}]
d = geography::Point(p)
      .STDistance(
         geography::Point(@$\phi$@, @$\lambda$@, 4326))
\end{lstlisting}

Here, SRID 4326 denotes the standard WGS-84 coordinate reference
system~\cite{nima2000wgs84}. Time complexity is $\mathcal{O}(n^2)$
where $n$ is the number
of bins; for typical routes of 15--20 stops this completes in under
100~ms, enabling real-time responsive optimization.

\begin{figure}[h]
  \centering
  \includegraphics[width=0.95\columnwidth]{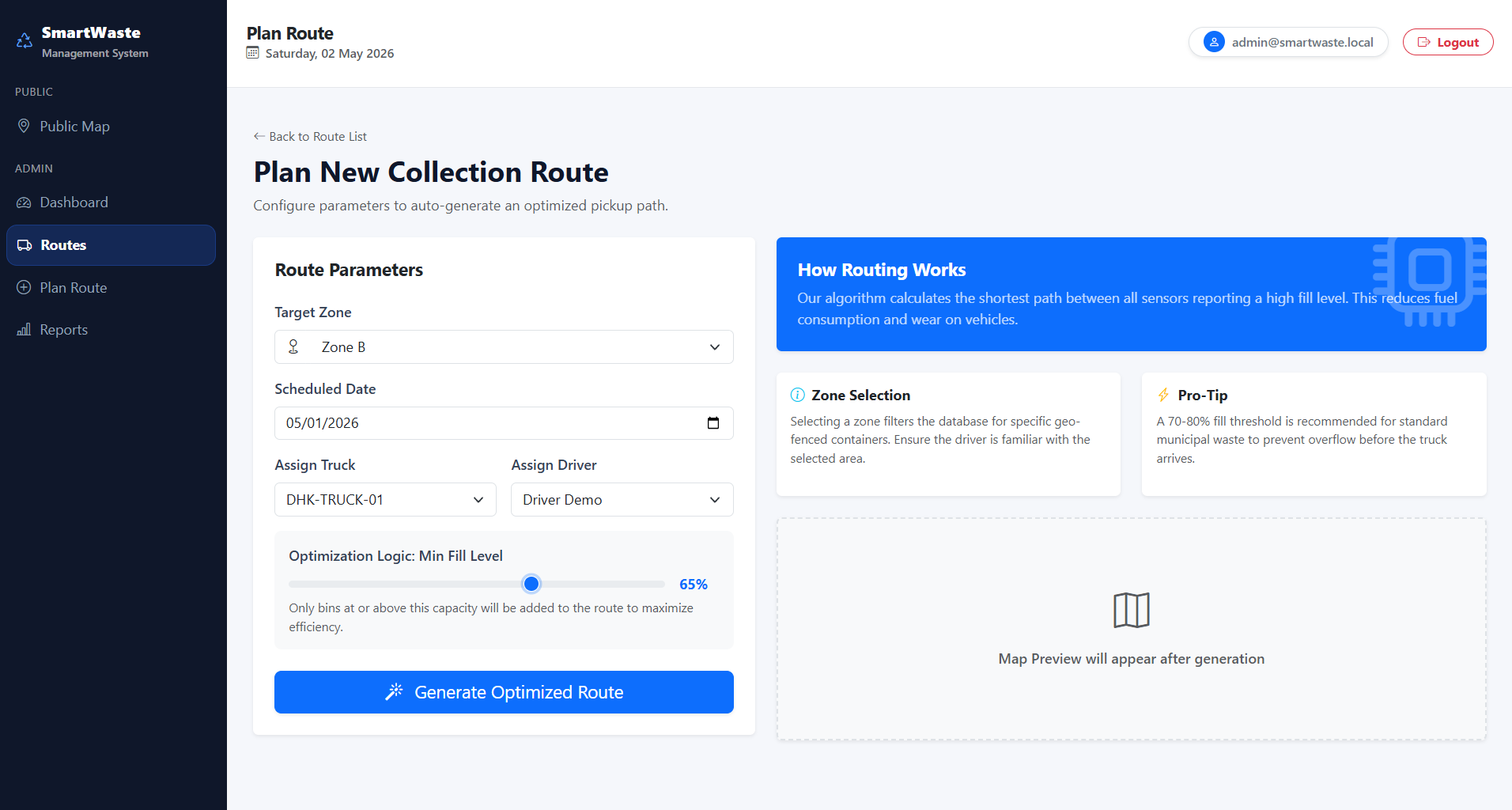}
  \Description{Screenshot of the Plan New Collection Route admin
  interface. The left panel contains route parameter inputs including
  a target zone dropdown set to Zone B, a scheduled date field, truck
  and driver assignment dropdowns, and a fill threshold slider set to
  65 percent. A Generate Optimized Route button appears at the bottom.
  The right panel explains how the routing algorithm works and shows
  a map preview placeholder.}
  \caption{Route planning interface: admin configures target zone,
  date, assigned vehicle and driver, and minimum fill threshold before
  triggering automated route generation.}
  \label{fig:plan_route}
\end{figure}

\begin{figure}[h]
  \centering
  \includegraphics[width=0.95\columnwidth]{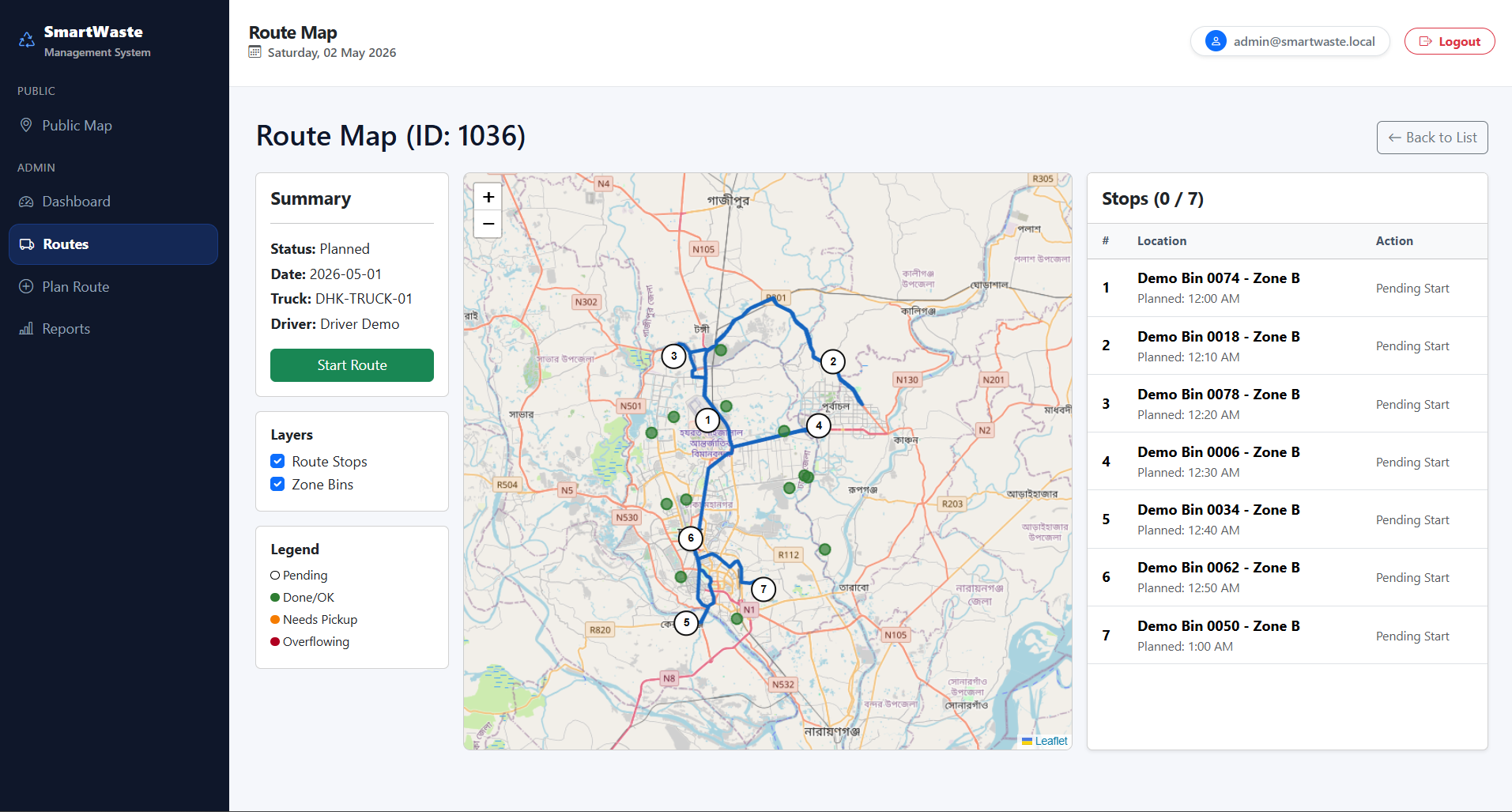}
  \Description{Screenshot of the admin route map view for Route ID
  1036. A Leaflet map of Dhaka shows a blue polyline connecting seven
  numbered stop markers in optimized sequence. A summary panel on the
  left shows status as Planned, date 2026-05-01, truck DHK-TRUCK-01,
  and driver Driver Demo, with a Start Route button. A stops list on
  the right shows all seven Demo Bin stops in Zone B with Pending Start
  status and planned times from 12:00 AM to 1:00 AM.}
  \caption{Optimized route map showing a 7-stop collection path with
  sequential numbered stops, route summary panel, and timestamped stop
  list.}
  \label{fig:admin_route}
\end{figure}

Figure~\ref{fig:plan_route} shows the route planning interface and
Figure~\ref{fig:admin_route} shows the resulting optimized route map.

\subsection{Driver Navigation and Collection Logging}

\begin{figure}[h]
  \centering
  \includegraphics[width=0.95\columnwidth]{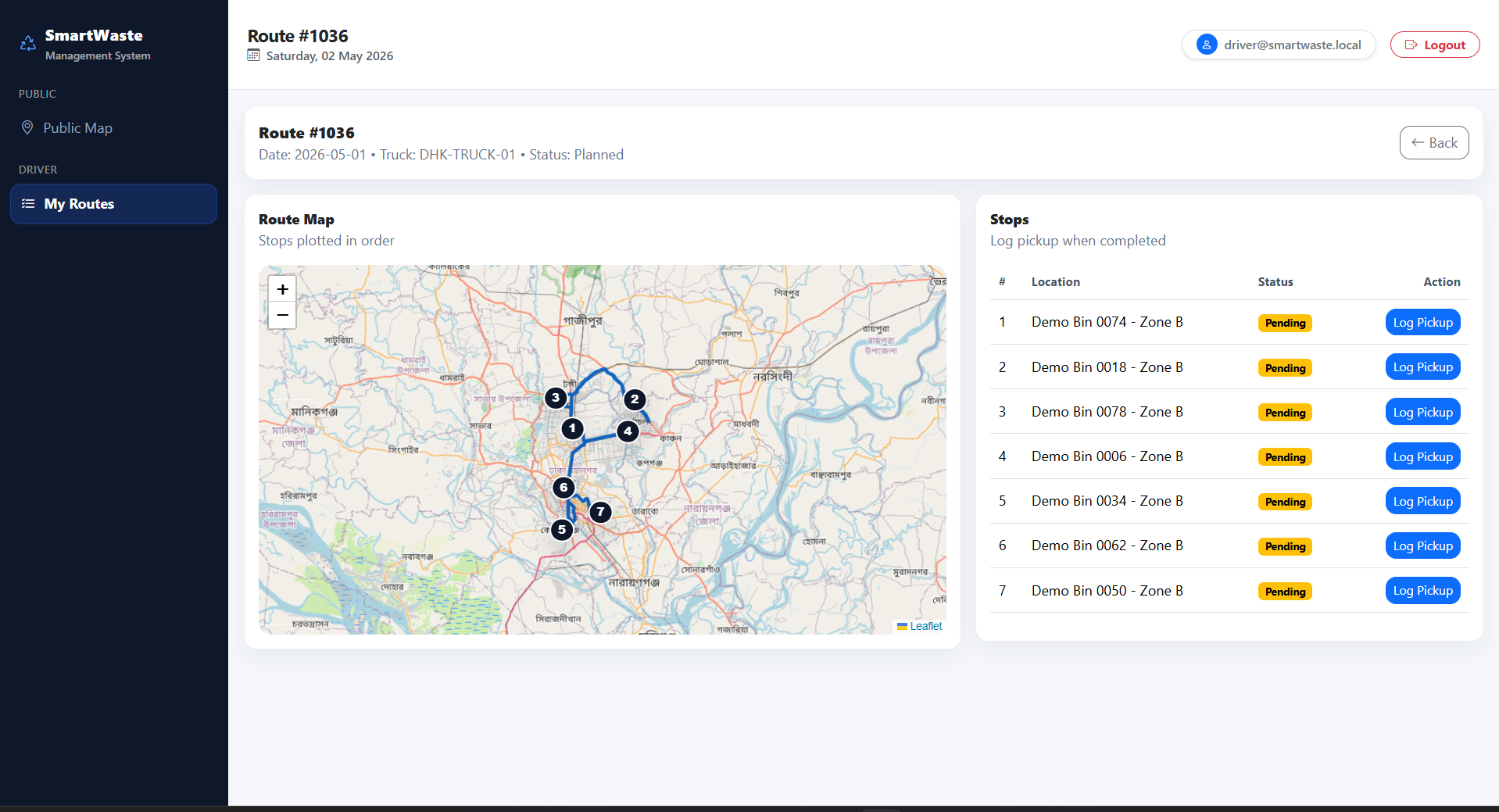}
  \Description{Screenshot of the driver route interface for Route
  1036. A Leaflet map shows seven numbered dark circular stop markers
  connected by a blue route line across Dhaka. On the right, a stops
  panel lists all seven Demo Bin locations in Zone B, each with a
  yellow Pending status badge and a blue Log Pickup action button.}
  \caption{Driver route interface showing 7 numbered stops in optimized
  sequence with individual ``Log Pickup'' action buttons.}
  \label{fig:driver_route}
\end{figure}

The driver interface (Figure~\ref{fig:driver_route}) displays the
optimized route as numbered stops on a Leaflet map. At each stop, the
driver logs: waste volume collected (litres), bin condition, and
optional photo. The system tracks completion progress and generates a
collection report (total waste, time, distance) upon route completion.

% ================================================================
\section{Experimental Results}
\label{sec:results}
% ================================================================

\subsection{Sensor Accuracy Across Calibrated Range}

To validate the configurable sensing approach, accuracy was tested at
five representative distance points within an experimentally calibrated
range ($D_{full} = 4.0$~cm, $D_{empty} = 73.0$~cm for this test bin).
The same formula (Eq.~\ref{eq:fill}) applies without modification for
any other calibrated range.

\begin{table}[h]
\small
\caption{Sensor Accuracy at Representative Fill Levels}
\label{tab:accuracy}
\begin{tabular}{p{1.2cm}p{1.4cm}p{1.1cm}p{0.9cm}}
\toprule
\textbf{Actual} & \textbf{Measured} & \textbf{Error} & \textbf{Fill} \\
\textbf{(cm)} & \textbf{(cm)} & \textbf{(cm)} & \textbf{(\%)} \\
\midrule
4.0  & 4.2  & 0.2 & 100\% \\
16.0 & 16.4 & 0.4 & 81\%  \\
34.0 & 34.8 & 0.8 & 55\%  \\
50.0 & 50.5 & 0.5 & 33\%  \\
73.0 & 73.0 & 0.0 & 0\%   \\
\midrule
\multicolumn{2}{l}{\textbf{Mean Absolute Error (MAE)}}
  & \textbf{0.38} & \\
\bottomrule
\end{tabular}
\end{table}

The MAE of 0.38~cm is within the rated hardware tolerance of the
HC-SR04 ultrasonic sensor ($\pm$3~mm)~\cite{hcsr04datasheet}, confirming
that measurement error is bounded by sensor physics rather than
software calibration. This translates to a fill percentage error of
less than 1\%, negligible for practical collection decisions. The
\texttt{constrain} clamping ensures stable 100\% and 0\% readings at
the calibrated boundaries regardless of minor sensor noise. The same
measurement accuracy is preserved for any other calibrated range since
the formula is geometry-independent.

\subsection{System Latency}

End-to-end latency was measured across 20 consecutive readings. Average
latency from sensor measurement to web platform dashboard update was
\textbf{5.3~seconds}, dominated by the fixed 5-second firmware polling
interval. Network transmission itself averaged under 300~ms on the
site-local Wi-Fi infrastructure, confirming that system response is
limited by design polling frequency rather than network overhead.

\subsection{Power Consumption}

The ESP8266 unit with active Wi-Fi and sensor polling draws 120--160~mA
at 3.3V, giving an average power consumption of $\approx$0.5W. The
integrated 5V/1W solar panel sustains continuous operation under normal
daylight and charges a backup battery for nighttime or overcast
operation.

\subsection{Mobile Web Interface}

\begin{figure}[h]
  \centering
  \includegraphics[width=0.60\columnwidth]{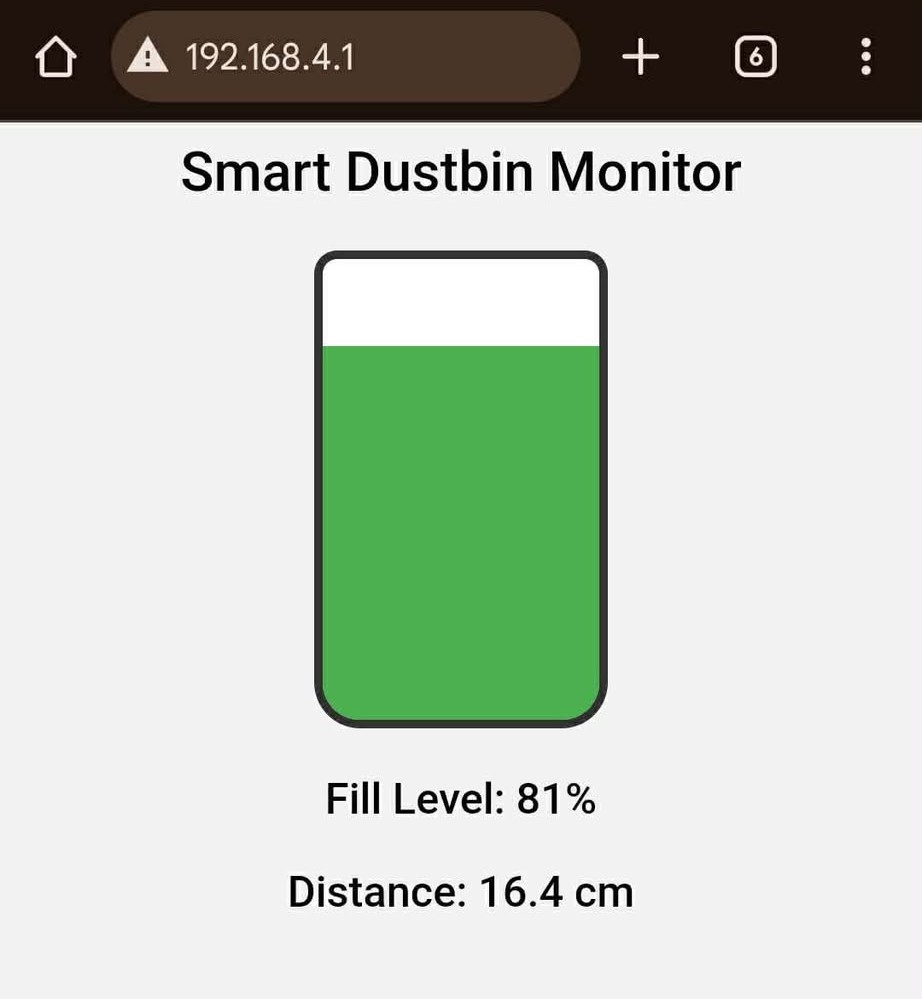}
  \Description{Mobile browser screenshot showing the ESP8266-hosted
  local web interface at IP address 192.168.4.1. The page is titled
  Smart Dustbin Monitor and displays a visual bin graphic filled
  approximately 81 percent with a green color. Below the graphic the
  text reads Fill Level: 81 percent and Distance: 16.4 cm.}
  \caption{ESP8266-hosted local web interface showing real-time fill
  level (81\%) and distance (16.4~cm) for on-site inspection.}
  \label{fig:demo}
\end{figure}

Figure~\ref{fig:demo} shows the local web interface hosted by the
ESP8266 at \texttt{192.168.4.1}, accessible from any device on the
bin's Wi-Fi access point. It displays fill level as a visual bar (81\%
in this reading) and the raw distance measurement (16.4~cm), allowing
on-site inspection without any external tools.

\subsection{Route Optimization Performance}

Route optimization was evaluated by creating collection routes for bins
distributed across a registered zone, with the nearest-neighbor
algorithm ordering stops using \texttt{STDistance()}-based geodesic
distances. Results are summarized in Table~\ref{tab:route}.

\begin{table}[h!]
\caption{Route Optimization Performance Metrics}
\label{tab:route}
\begin{tabular}{lc}
\toprule
\textbf{Metric} & \textbf{Value} \\
\midrule
Algorithm execution time (avg.)      & $<$100~ms \\
Distance function used               & \texttt{STDistance()} (WGS-84) \\
Coordinate reference system          & SRID 4326 \\
Admin-to-driver route delivery       & $<$1.5~s \\
Bins filtered by fill threshold      & $\geq$70\% (configurable) \\
Stop ordering basis                  & Geodesic proximity \\
\bottomrule
\end{tabular}
\end{table}

The sub-100~ms computation time enables responsive real-time
optimization, and the operator can immediately reassign or regenerate
routes if collection priorities change. Using \texttt{STDistance()}
rather than a planar approximation ensures that stop ordering remains
accurate across all urban-scale deployments regardless of geographic
location.

\subsection{Comparison with Existing Smart Waste Management Systems}

Table~\ref{tab:comparison} compares CleanCity-BinSense with
representative systems. Compared with previous work, the proposed
platform combines configurable per-bin sensing, embedded route
optimization, role-based management, and self-hosted deployment within
a single architecture. Unlike cloud-dependent or GIS-assisted systems,
it achieves \textbf{0.38~cm} sensing MAE, \textbf{5.3~s} end-to-end
latency, and \textbf{sub-100~ms} route generation on a complete
hardware-software prototype using commercially available components,
demonstrating its practicality for municipal deployment.

\subsection{Network Reliability}

During experimental evaluation, data transmission over the site-local
Wi-Fi network remained stable, with successful delivery of sensor
readings throughout repeated testing. The observed average end-to-end
latency of 5.3~seconds was primarily determined by the fixed sensing
interval rather than communication delays, while network transmission
typically required less than 300~ms. These results indicate that,
under reliable local network conditions, communication overhead has
minimal impact on overall system responsiveness.

% ================================================================
\section{Discussion}
% ================================================================

The experimental results demonstrate that CleanCity-BinSense enables
practical real-time waste monitoring and collection optimization for
resource-constrained environments. The configurable sensing approach
achieved reliable measurements across the calibrated range, while the
system maintained stable communication and generated optimized routes
in sub-100~ms. Together, these results validate the proposed hardware,
software, and optimization framework as a scalable solution for
heterogeneous urban waste collection.

\textbf{Scalability:} The most significant practical advantage of
CleanCity-BinSense is its configurable sensing range, which enables
straightforward scaling across heterogeneous urban waste networks. In
real deployments, bins of varying depths (e.g., 40~cm, 80~cm, and
120~cm) coexist on the same collection network. CleanCity-BinSense uses a single firmware with two per-bin calibration parameters
set during installation. Combined with the modular
architecture, each sensor node operates independently while
communicating with a centralized server. This enables incremental
expansion from small pilot deployments to city-wide implementations
without requiring hardware or firmware modifications.

\textbf{Security:} The current prototype operates within a trusted
site-local network where communication occurs over HTTP between
authenticated hardware and the server. User access to the management
platform is protected through role-based authentication, restricting
administrative, operational, and driver functions according to user
privileges. For large-scale municipal deployment over public networks,
additional security measures such as HTTPS encryption, device
authentication, API tokens, secure credential management, and regular
software updates should be incorporated to protect sensor data and
prevent unauthorized access.

\textbf{Limitations:} The nearest-neighbor heuristic provides practical
routes but does not guarantee global optimality. For larger route sets
($n > 50$), algorithms such as Christofides or genetic algorithms may
improve performance, while pre-computed distance matrices could further
reduce planning latency. Large-scale deployment also requires one-time
sensor calibration for different bin geometries and periodic
maintenance to mitigate environmental effects. The current system
relies on site-local Wi-Fi; deployments without existing network
infrastructure would require 4G or NB-IoT connectivity. GPS coordinates
are manually configured during installation, with autonomous location
reporting left for future work.

\textbf{Long-Term Operation and Energy Considerations:}
The solar-powered ESP8266 prototype consumed approximately 0.5~W during
normal operation, with the integrated charging system supporting
continuous daytime operation and nighttime battery use. Future work will
evaluate durability, battery longevity, sensor stability, and
communication reliability through extended field deployments.

% ================================================================
\section{Conclusion}
\label{sec:conclusion}
% ================================================================

This paper presented CleanCity-BinSense, a complete IoT-enabled smart
waste management system featuring a solar-powered sensor unit with a
configurable sensing range for diverse bin geometries, a role-based web
platform, and automated nearest-neighbor route optimization using SQL
Server's \texttt{STDistance()} function. Experimental results
demonstrated a sensor MAE of 0.38~cm, within the ultrasonic sensor's
rated tolerance, 5.3~s end-to-end latency dominated by the sensing
interval, and sub-100~ms route generation, demonstrating the system's
suitability for deployment in resource-constrained urban environments
such as Dhaka, Bangladesh. Future work includes (1) machine learning
for predictive fill scheduling, (2) 4G/NB-IoT connectivity, (3) onboard
GPS modules, (4) advanced route optimization (Christofides or genetic
algorithms), (5) gas sensor-based waste classification, and (6) a
100-bin multi-zone pilot deployment.

% ================================================================
% REFERENCES
% ================================================================

\end{document}